# UNDERSTANDING THE INTERPLAY BETWEEN BOUNDARY RESOURCES AND GOVERNANCE PRACTICES IN INFLUENCING ECOSYSTEM VALUE CO-CREATION FOR DIGITAL PLATFORMS: A CASE FROM THE GLOBAL SOUTH

Elijah Chirwa, University of Sheffield, echirwa1@sheffield.ac.uk

Pamela Abbott, University of Sheffield, p.y.abbott@sheffield.ac.uk

Jonathan Foster, University of Sheffield, j.j.foster@sheffield.ac.uk

**Abstract:** Despite their potentially transformative impact, few studies have investigated how commercially-driven digital platforms such as mobile money facilitate ecosystem value co-creation in the global South. Consequently, using a boundary resources model and platform governance approaches, this paper aims to examine how a payment platform facilitates ecosystem value co-creation with third party actors. An in-depth qualitative inquiry was adopted for the study using an embedded single-case design. The results show that although ecosystem value co-creation was enabled by the introduction of boundary resources tools, some platform governance practices hindered some complementors from harnessing the affordances of the platforms. These constraints include lack of visibility of the boundary resources and transparency challenges in the acceptance criteria. We thus argue that platform generativity on its own is not sufficient to support ecosystem value co-creation, but requires appropriate platform governance to deal with behavioural complexity of ecosystem actors by using optimal control mechanisms.

**Keywords:** digital platforms, value co-creation, boundary resources, platform governance, digital affordances, generativity.

## 1. INTRODUCTION

In the new global economy, digital platforms are becoming pervasive and powerful, primarily as they enable transactions between different groups of actors and facilitate innovation of products and services (Cusumano, Gawer, & Yoffie, 2019). Recent evidence suggests that the advent of digital platforms has democratized innovation, shifting it from being a standalone firm-centric activity to open and distributed ecosystem-based value co-creating practices, extending beyond a single firm's locus (Chesbrough, 2003; von Hippel, 2005; Yoo, Henfridsson, & Lyytinen, 2010). The impact of this phenomenon is also being experienced in the global South as digital-based platforms enable ecosystems that facilitate value co-creation amongst a diverse set of actors (David-West & Evans, 2015).

Koskinen et al. (2019) argue that although many digital platforms first emerged in the global North, they have also gained wide usage in the global South due to the proliferation of mobile devices and ubiquitous connectivity. Despite significant challenges encountered during the creation and scaling of digital platforms in the global South, which include deficient infrastructure, market constraints, and lack of consumer trust and knowledge, these platforms possess the potential to address some of the social and economic challenges faced by the people at the bottom of the income pyramid (David-West & Evans, 2015; Koskinen, Bonina, & Eaton, 2019; Praceus, 2014). Additionally, Heeks et al. (2020) argue that digital platforms can potentially address the perennial issue of institutional voids that have caused market constraints and hindered the successful implementation of previous technology-led initiatives in the global South. Mobile money platforms are a typical example of a digital platform offering transformative financial services both as a transaction and innovation platform in the global South.

However, despite their potentially transformative impact, there is a paucity of studies that investigate how commercially-driven digital platforms such as mobile money facilitate ecosystem value co-creation in the global South. Existing research conducted on how platform owners stimulate third party innovation on platform ecosystems in the global South has focused on non-commercial digital health platforms, which present different operating dynamics compared to profit-driven digital payment platforms (Msiska, Nielsen, & Kaasboll, 2019). This lack of research further provides an opportunity to explore how the dynamics of platform-based service innovations such as mobile money are integrated and used to facilitate complementor value co-creation. These third party innovations have potential to address contextual challenges specific to these low-resource settings. Although third-party developers may have the capacity to co-create value on such digital platforms that can address context-specific challenges in the global South, they face numerous governance related challenges that hinder their participation in platform ecosystems.

Therefore, using the case of mobile money in Malawi, this paper aims to investigate how commercially driven digital platforms facilitate ecosystem value co-creation with complementors. More specifically, the study addresses these questions: (a) what boundary resources are provided by the platform owner to support ecosystem value co-creation? (b) what governance practices have been implemented as control mechanisms to secure the digital platform? (c) how does the interplay between boundary resources and governance practices influence ecosystem value co-creation for digital platforms in a global South setting? This research adopts the boundary resources model as the conceptual framework to analyse ecosystem value co-creation since this model has been shown to play a vital role in understanding platform owners' governance approaches, which ultimately influence complementor outcomes (Eaton, Elaluf-Calderwood, Sørensen, & Yoo, 2015; Ghazawneh & Henfridsson, 2013; Huber, Kude, & Dibbern, 2017).

# 2. THEORETICAL BACKGROUND

## 2.1. Digital platforms as value co-creating ecosystems

Cusumano et al. (2019) divide digital platforms into three types, firstly innovation platforms which are characterised as those that provide the building blocks for innovation and enable recombination of heterogeneous functionality on the platform from ecosystem actors. Secondly, there are transaction platforms largely comprising those that play an intermediary role or provide marketplaces for interaction between various sides of a market. The third type of platform comprise those that exhibit and support both innovation and transaction characteristics. A common feature amongst all these different types of digital platforms is that they are all underpinned by digital infrastructures that facilitate innovation between the platform owner and third-party developers (complementors). Tilson et al. (2010, p. 748) describe digital infrastructures as "basic information technologies and organisational structures, along with the related services and facilities necessary for an enterprise or industry to function". Thus, digital platforms represent a socio-technical collection comprising digital infrastructures and associated organisational agents, processes, and standards that enable different actors to orchestrate their service and content needs (Constantinides, Henfridsson, & Parker, 2018; De Reuver, Sørensen, & Basole, 2018; Tilson, Lyytinen, & Sørensen, 2010). Thus, digital infrastructures inherently possess properties that facilitate generative value co-creation which can be harnessed by the wider community in the ecosystem. Generativity refers to the "capacity to produce unprompted change driven by large, varied and uncoordinated audiences" (Zittrain 2006, p.1980).

Digital platforms provide a functional utility for the ecosystem actors to potentially undertake value co-creation afforded by the properties of the underlying digital infrastructure (Autio & Thomas, 2019). However, the value from this functional utility is usually undefined and unknown, and can thus only be realised if the ecosystem generativity is harnessed. Consequently, Bonina and Eaton (2020) argue that the platform owner plays a vital role in nurturing the platform ecosystem's growth by enabling innovation activities that facilitate value co-creation and by preventing actions that drain value away from the platform. Bianco et al. (2014) point out that platform ecosystems' success depends highly on the diversity and value of end-user products and services. This is evident in the case of the flagship M-pesa, a thriving mobile money digital platform from the global South, where value co-creation enabled third-party developers to extend the functionality of the platform. Research shows that the platform owner for M-pesa played a significant role in enabling value co-creation with complementors and end-users to address context-specific needs, which ultimately contributed to the platform's success (Kendall, Maurer, Machoka, & Veniard, 2011; Markus & Nan, 2020; Mwiti, 2015).

In spite of the clear evidence of the value co-created by the diverse actors interacting on the digital platform and the potential success it may bring, the literature also shows unique organisational challenges and tensions that exist between the platform actors in the ecosystem. For example, Tilson et al. (2010) refer to the governance challenge of balancing different interests of ecosystem participants between platforms' stability and flexibility as it determines the realization of ecosystem value co-creation. This tension is one of the critical challenges in developing digital platforms' innovations and entails a wider role for platform owners that involves platform governance. Previous studies have defined platform governance as the fundamental decisions undertaken by the platform owner in relation to complementors, including the ownership of the platform and the interactions with the ecosystem of complementors (Boudreau, 2012; Gawer, 2014; Wareham, Fox, & Giner, 2014).

Although research has ascertained platform ecosystems' rising importance in facilitating value co-creation, platform owners may still encounter challenges in addressing context-specific needs of distant and unknown end-user communities due to lack of familiarity with them (Bosch, 2009; Henfridsson & Lindgren, 2010). The challenges are further exacerbated by the contrasting requirements of different groups of end-users that might be beyond the focus or core expertise of the platform owner (Boudreau, 2010; Ghazawneh & Henfridsson, 2013). The literature on consumer innovation in BoP contexts further highlights the challenges that service providers face, such as lack of knowledge concerning the lives, needs information, and the preferences of the poor, which calls for local embeddedness into innovation ecosystems that support co-creation of value (Cañeque & Hart, 2017; Praceus, 2014; Viswanathan & Sridharan, 2012). Accordingly, extending platform functionality to complementors close to the end-user contexts is becoming increasingly appealing to platform owners as it enables third-party developers to address context-specific needs (Ghazawneh & Henfridsson, 2013; Msiska, 2018).

### 2.2. Boundary resources model: governance mechanisms for ecosystem value co-creation

Foerderer et al. (2019) argue that examining the complex and dynamic socio-technical interactions at the boundary between platform owners and complementors offers an opportunity to gain more insights into platform governance and ecosystem related tension. In resolving the ecosystem tension, the platform owner undertakes a delicate balancing act between opening up the platform functionalities to complementors for value co-creation and maintaining optimal control over the platform ecosystem through boundary resources and boundary-spanning activities (Eaton et al., 2015; Ghazawneh & Henfridsson, 2013; Huber et al., 2017). Boundary resources are defined as "the software tools and regulations that serve as the interface for the arm's length relationship between the platform owner and the application developer" (Ghazawneh and Henfriddson, 2013, p.174). Boundary resources provide governance mechanisms that help to manage the socio-technical interactions between the actors, and serve as the interface between platform owners and complementors in the value co-creation process (Bianco, Myllarniemi, Komssi, & Raatikainen, 2014; Eaton et al., 2015). Consequently, since boundary resources facilitate access to core platform services and fuel generative value co-creation in digital ecosystems, they also play an important role in shaping complementor outcomes. Therefore, this paper adopts the boundary resources model to gain insights into how platform owners facilitate ecosystem value co-creation as they manage ecosystem governance through enforcing control and enabling platform generativity.

The boundary resource model provides a tool to analyse two vital roles in platform governance, which are resourcing and securing the platform, as depicted in Figure 1. Examples of tools for resourcing include application programme interfaces (APIs) or software development kits (SDK), both of which enable developers to access the platform's core resources (Ghazawneh & Henfridsson, 2013). The second governance role is to enable the platform owner to secure platform control and maintain its integrity by providing appropriate rules and regulations that ensure the overall quality is not compromised and remains in line with the platform’s goals (Boudreau, 2010). A typical example of a securing role would include implementing a set of guidelines, rules or activities for platform complementors. The rules and tools provided by boundary resources assist in understanding platform owner’s control and complementor’s contribution towards innovation.

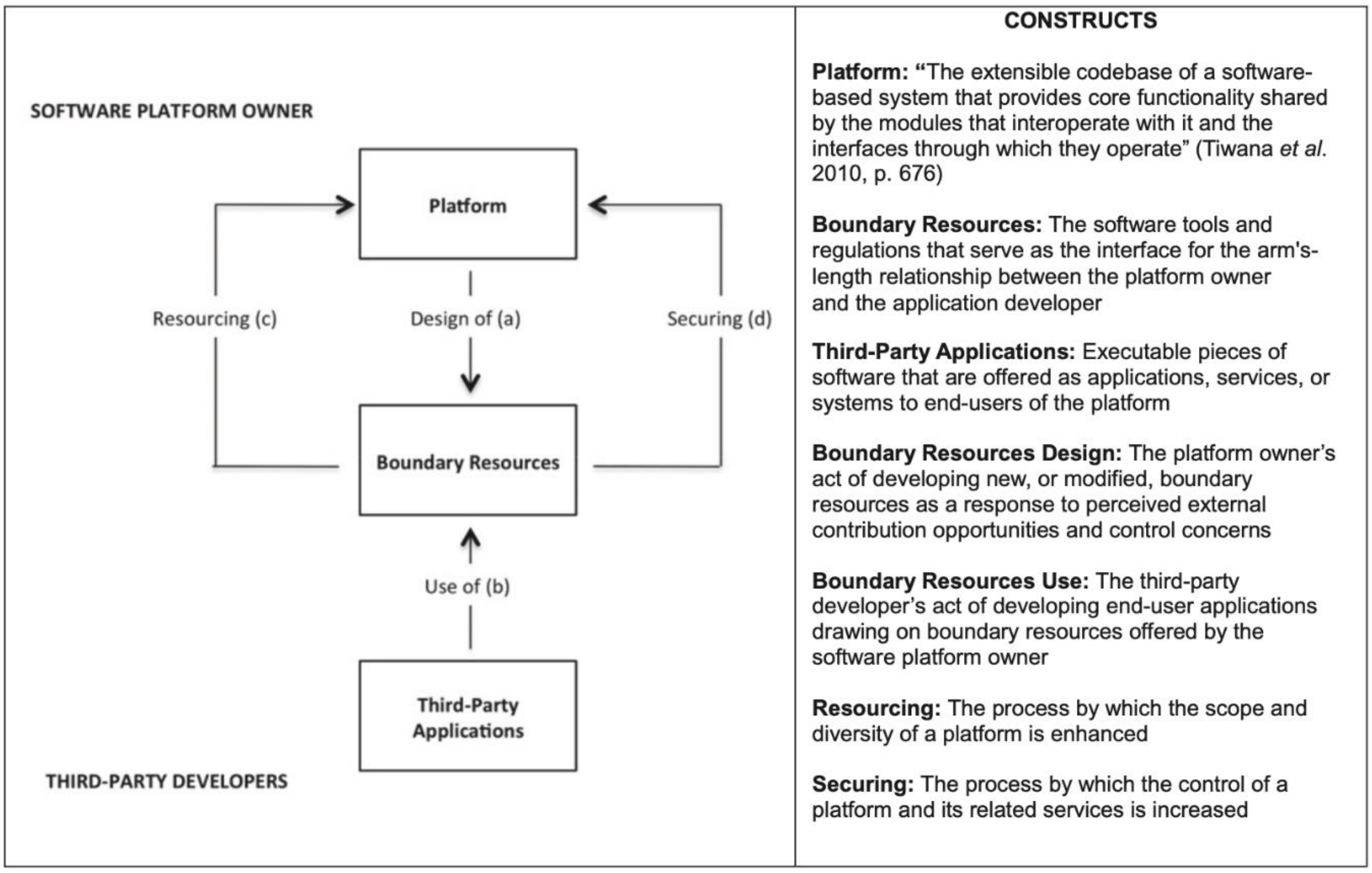

Figure 1: Boundary resources model Source: Ghazawneh and Henfridsson (2013)

# 3. METHODOLOGY

## 3.1. Research design

In order to investigate how a mobile money digital platform facilitates ecosystem value co-creation within its surrounding context, an in-depth qualitative inquiry was adopted using an embedded single-case design. This approach was chosen due to its suitability to understand the dynamics of a novel phenomenon and its context in a single setting (Yin, 2018). The embedded single-case approach was utilised with the aim of incorporating multiple units of analysis at different levels of the case, with the platform ecosystem offering the overarching unit of analysis whilst the boundary resources and complementors provided the primary subunit for the study. These varied embedded subunits at different levels of the study avail an opportunity for detailed analysis that enhances insights derived from the research of the phenomenon. The case study strategy resonates well with the aims of the study as it allows us to understand the phenomenon in its real-life setting and gain deep insights into the development and emergence of value co-creating practices for the mobile money digital platform (Benbasat, Goldstein, & Mead, 1987; Saunders, Lewis, & Thornhill, 2018).

### 3.2. Case description – Mobile money digital platform

This case study is based on the mobile money digital platform in Malawi and its digital innovation ecosystem, which is referred to here as Alpha. Alpha platform commenced provision of mobile money services in early 2012, focusing on facilitating exchange between cash and electronic value through mobile phones for various entities on its transaction platform. Between 2012-2016 Alpha transitioned to a start-up stage, where the platform offered some basic first-generation services and products (UNCDF, 2018). From 2016, the platform commenced its main expansion and consolidation phase, which has been characterised by the deployment of more services and facilitation of partnerships due to its increased technical capabilities to deliver additional value-adding services for the platform. This study focuses largely on this latter phase as the platform owner is able to allow and foster value co-creation with third-party firms as a way of extending the platform functionalities. Furthermore, the increasing need for a wide array of actors to integrate into the mobile money ecosystem as well as the generative affordances that the platform offers in addressing varied user needs in Malawi make this an interesting and appropriate case for this research study.

### 3.3. Sample and sampling techniques

The participants for the study were selected from the ecosystem actors constituting the mobile money innovation ecosystem, which comprised the platform owner, the financial regulator, banks, civil society organisations and third-party developers, who included complementors. On average, two interviewees were selected from each participant category through purposeful sampling. The participants also included a group of ecosystem actors who have either developed complementary services or attempted to develop third-party services for the platform. A total of twelve semi-structured interviews were conducted for this research, five of which were face to face and the remaining over Skype due to Covid-related constraints.

Informed and voluntary consent was obtained from the research participants as part of the ethical considerations for the study. This involved full disclosure to all participants of the purpose of the study and adequate details regarding how to participate, decline or exit from the study at any point during the course of the research.

### 3.4. Data collection and analysis

The data collection was conducted primarily through semi-structured interviews as they afford the researcher an opportunity to gain first-hand views as well as room for improvisation and probing of the studied phenomenon (Hein et al., 2019; Walsham, 1995). The areas of focus for the research included aspects of how the platform owner facilitated resource integration with complementors to achieve value co-creation, understanding in detail the governance mechanisms employed by the platform owner in supporting generative innovation, and examining contextual issues surrounding ecosystem value co-creation.

An abductive approach to theory building was used and included iterations between data and theory to help uncover deep insights into our research study (Dubois & Gadde, 2002). The theoretical constructs were derived from the literature on platform governance and ecosystem value co-creation with concepts from the boundary resources model guiding the initial coding process (Ghazawneh & Henfridsson, 2013, 2015). The key focus of the analysis was to explore the phenomenon of ecosystem value co-creation and the role of governance mechanisms in supporting generative innovation with the goal of identifying key themes that could explain the noted patterns from the data and match these to the theoretical concepts.

# 4. FINDINGS

In this section, we present the results on how the platform owner has facilitated ecosystem value co-creation and the role of governance mechanisms in supporting generative innovation.

## 4.1. Resourcing platforms to enable ecosystem value co-creation

During its early stages, Alpha provided mobile money services that were largely developed within the firm. However, this changed in 2016, when the platform owner undertook a technological upgrade that enabled a transformation of its digital platform to facilitate collaboration and integration with external third-party actors. The upgrade enabled the platform to provide an API that granted access to external third-party actors to develop services on top of the mobile money service, as explained by the platform owner:

> *"Initially, the system was able to accommodate the original requirements of sending and receiving electronic money which were the original needs of a mobile money service. However, gradually we realised that the system was not flexible enough to extend its functionalities to third-parties and yet the mobile money service needs kept growing with more integration and development requests coming from other firms... and thus it became obvious that we needed to upgrade our system"[Res1].*

These newfound digital platform capabilities were confirmed by several participants:

> *"…the upgrade allowed our firm to offer API functionalities to some external third-party companies, who were then able to integrate to our platform and develop new innovations in line with our rules" [Res3]*

> *"Our firm provides micro loan targeting the unbanked based on their phone usage, and the presence of the API enabled us to integrate to the platform and provide the service to any eligible phone user" [Res9]*

The ability to integrate seamlessly with third-party entities through the API enabled complementors such as a micro loan firm to provide new services to a wider group of end-users by harnessing the potentially transformative capabilities of the platform. Indeed, the results show that various services were developed by third-party actors on the platform ecosystem:

> *"So the micro loan product was developed by combining our specialised knowledge and development resources with some core data from the platform owner and other capabilities of the digital platform to come up with a credit rating score for mobile phone users. Then the service runs an algorithm that will determine the loan thresholds for various users. We also partnered a bank which then provides the resources for these loans." [Res9]*

> *"We developed a mobile payment product in collaboration with the mobile money platform owner with whom we partnered to come up with a solution using their platform capabilities and our own expertise to come up with an innovation that enabled farmers use cashless payments to procure agricultural seeds at the same time the platform provided market access to the seed company." [Res6]*

It was also noted that the availability of the API enabled technical integration between the ecosystem actors. Several interviewees highlighted the importance of the API as a tool to achieve integration:

> *"The API allowed us to integrate with various firms such as banks, as they used the platform as one of their service delivery channels; innovators that came up with services which operate exclusively on mobile money platforms; and software developers that offered mobile money integration services to other firms"* [Res2].

> *"In the absence of the API from Alpha, our firm would not have been able to use mobile money as a service delivery channel for our products"* [Res12]

Upon being approved to integrate to the platform, Alpha provides standardized interfaces to the external actor to facilitate development of additional services on the mobile platform. Describing this stage further, one of the participants commented:

> *"We started the integration work by being provided the documentation of the API and being informed of the other processes and technical tasks that we needed to undertake to achieve the integration. It was clear in terms of what we were allowed or not allowed to do from accessing the API itself and the documentation that was provided." [Res9]*

However, there are also divergent views among the complementors who are already integrated to the platform as to how the platform owner supports ecosystem value co-creation as evidenced by these two participants:

> *"At some point, we experienced challenges with the standard API as presented on the platform to integrate with our banking application as it could not meet some of our needs and therefore we had to put in a request to the platform owner who accepted our request and went ahead to make the necessary changes which worked to our satisfaction"* [Res12]

> *"We have had challenges to have one of our requests with the integration since we went live over a year ago and we needed changes on the API but to-date nothing has materialised"* [Res5]

The results show that the platform owner lacks consistency in addressing issues from the complementors. Additionally, as part of extending the platform's diversity and providing a bridge to startup firms which face challenges to meet platform owner's conditions, it was noted that an internationally funded firm had installed a hub solution that served as a single entry point to the platform for some third-party actors:

> *"We were approached by an international organisation requesting to be allowed to install a hub solution that would serve as an entry point to the mobile platform for smaller firms. The organisation is targeting small entrepreneurs who provide critical utilities such as water and power to off-grid populations that pay small amounts and may thus require to use mobile money service for such payments"* [Res3]

The provision of a hub solution is aimed at enabling smaller firms to participate on the Alpha platform ecosystem and bring their innovations to a wider audience. The target group are those entrepreneurs who that may not have resources required by the platform owner.

### 4.2. Supporting innovation by securing the platform

In line with its governance approach to securing the platform, Alpha has implemented a number of processes as part of its control mechanisms. The processes impact the decision-making rights of the platform actors and provide control mechanisms for the platform owner such as gatekeeping and enforcement of regulations. It was noted that the platform owner makes all key decisions pertaining to the platform and uses tools such as contract agreements and platform rules to provide regulation-based governance in addition to its technology-based controls. Several interviewees confirmed this arrangement:

> *"The responsibility to accept or reject a third to integrate to the platform rests with us as the platform owner and is based on several factors" [Res1*]

> *"Upon being accepted to integrate to the mobile money platform, we were requested to sign a contract agreement with the platform owner" [Res6]*

The decision-making process and the regulatory tools enable the platform owner to use non-technical governance tools as a means of managing innovation on the platform. Additionally, to ensure controls in accessing the platform, the owner evaluates the business viability before any third-party actors are granted access to the platform. One research informant explained the process as follows:

> "*A rigorous assessment of the business case in line with platform owner's strategy whereby they decline the services if they are not sure of its viability or not in tune with their overall strategy" [Res6]*

It was stated that these measures by the platform owner were aimed at controlling access by ensuring that the quality of the services developed for the platform are in tune with the owner's overall strategy. In line with this control mechanism, there seems to be a deliberate approach for limited communication on the availability of the API:

> *"We have not yet openly published the API on our website" [Res3]*

> *"I am surprised to hear that Alpha platform has an API that can enable small firms like ours to bring innovations to the platform. I approached them some few years back and they indicated that they don't have an API for third-parties and haven't tried again since that time"[Res11]*

This lack of adequate communication for the API to external parties led to a number of firms being unaware of their existence. Additionally, a number of respondents from third-party firms that were aware of the API availability had been rejected to develop services on the platform and they highlighted their experiences:

> *"We tried to integrate our village savings and loan association application targeting the unbanked to the mobile money platform but have faced numerous challenges mainly due to the business viability disagreements, prohibitive costs and lack of trust in our firm"* [Res10]

> *"We were declined access to Alpha mobile platform in a manner that lacked a clear and consistent approach, however are now just trying to find alternative ways of enabling payment functionalities for our service which we know the mobile money platform could have provided the best solution"*[Res7]

> "*We approached the platform owner on several occasions to integrate our transportation application which we wanted to integrate with the mobile money platform, however on numerous occasions we were informed that their API is not available despite being available on some banking websites. We are now pursuing other means to achieve the same goal*" [Res11]

These statements reveal that varied sentiments exist amongst third-party firms as regards the API such that even some of the firms that were aware of its availability and had attempted to gain access were nevertheless declined access due to various reasons. Furthermore, it shows that Alpha puts some restrictions as regards who can participate on their platform and the criteria was based on the platform owner's goals and objectives.

# 5. DISCUSSION

The results have demonstrated that Alpha's design of platform boundary resources enabled the introduction of the API as a software tool and heralded a new era for the platform as it opened up opportunities for the owner to collaborate with third-party actors in co-creating value beyond the boundaries of the firm. The API enabled the platform to respond to external contribution opportunities and thus offer digital affordances that extend the scope and diversity of the platform functionalities. This is consistent with the intellectual account provided by the boundary resources model (Eaton et al., 2015; Ghazawneh & Henfridsson, 2013). The platform owner offers digital affordances through the boundary resources that in turn facilitate third parties in developing services on the digital platform. This finding confirms that boundary resource tools such as APIs are indispensable in extending platform's functionality as they stimulate external contribution opportunities by allowing complementors the capability to provide additional applications for the platform ecosystem.

This case study has also shown how the platform owner, through the provision of APIs, enables a third-party actor, such as the macro loan provider, to use boundary resources tools to integrate and develop complementary services by harnessing the digital affordances of the platform, and thus enabling ecosystem generativity. The result confirms that platform value co-creating mechanisms are dependent on the provision of digital affordances, which enable platform generativity (Nambisan, Song, Lyytinen, & Majchrzak, 2017; Yoo, Boland, Lyytinen, & Majchrzak, 2012). The result explains how the Alpha digital platform ecosystem as an affordance platform provided a functionality role through boundary resources and orchestrated generativity, which allowed value-creating services to emerge. These value co-creation instances allowed the enhancement of the scope and diversity of the Alpha mobile money platform.

However, despite some complementors being able to harness the generativity of the platform's affordances to co-create value, the results also show some challenges faced by some complementors due to a "deliberate" lack of visibility of the boundary resources and transparency challenges in acceptance criteria. In some instances, third-party actors were either denied access to the platform through a non-transparent process or were not even aware of the availability of boundary resources. These actions curtail the development of further value co-creating services as third party actors with potential generative capacities are denied the opportunity to co-create value. These observations match those expressed by several other authors that to successfully build platform-centric ecosystems, the platform owner must shift the design capabilities to external actors so as to enhance their generative capacity (Avital & Te'Eni, 2009; Msiska, 2018; Von Hippel & Katz, 2002). This reinforces the point that generativity is a potential passive capability that needs to be activated; otherwise, it is not useful. Furthermore, this lack of transparency seems to support observations from some third parties that there is an inconsistent application of the control mechanisms against smaller firms. In contrast, Tiwana (2013) identified transparency and fairness as some of the critical elements in designing effective platform controlling mechanisms. The seeming lack of fairness and transparency in our study might deter other third-party innovators from joining the platform and discourage others from remaining on the platform, thus impacting generativity. Therefore, although Alpha platform provides generative capability, the lack of visibility and transparency of boundary resources stops complementors from offering new resources, knowledge and capabilities that would harness platform affordances.

The findings reveal that the platform owner designed the boundary resource rules such as contract agreements, platform acceptance criteria and API guidelines as non-technical governance control mechanisms for exerting its authority and decision making responsibility over the platform. For example, the use of a business case assessment as a predefined condition to be allowed on to the platform ecosystem denotes a gatekeeping role being used as a control mechanism for accommodating third-party actors to use the boundary resources. These observations confirm the vital role played by the platform owner in securing the platform by employing control mechanisms over heterogeneous actors to enforce rules in line with its interests (Constantinides et al., 2018; Ghazawneh & Henfridsson, 2013; Van de Ven, 2005). The use of these various control mechanisms assisted Alpha to ensure third-parties develop services that are aligned with its interests.

The introduction of a new boundary resource tool as a "hub solution" by an international organisation represents the emergence of distributed governance that allows smaller entrepreneurs circumvent some controls exerted by the platform owner. This hints at some underlying challenges facing this group of startup firms which need to be overcome to open up the platform to a wider set of actors and innovations. This is supported by the observation that the emergence of new boundary resources can be triggered by either the owner's perceived external contribution opportunities or by external third-party use of the boundary resources (Ghazawneh & Henfridsson, 2013). The development of the hub solution seems to emanate from the need to support marginalized third party actors to innovate for the platform as they address context specific needs.

Overall, the underlying assumption in digital platform ecosystems is that generativity drives ecosystem value co-creation. However, this ecosystem value co-creation is determined by how platform owners overcome the tension between supporting generative capacity of autonomous individual actors that provide unpredictable innovative inputs, thus requiring some control, and the logic of technological flexibility that requires stability (Eaton et al., 2015; Huber et al., 2017; Wareham et al., 2014). Our findings support these assertions on how digital affordances enabled generative value co-creation and also how governance approaches using control mechanisms attempted to address behavioural complexity but in the process constrained generativity. It is evident that most of these constraints are coming from the control mechanisms and that 'circumventing' and allowing the platform to create opportunities for further value co-creation has been necessary by introducing different types of governance approaches such as the hub solution.

The results also suggest that, as a commercially driven platform, Alpha focuses on economic value, cost-benefit trade-offs and producer-centric value co-creation. As a result, due to its governance approach of centralised authority and responsibility, it runs the risk of under-appreciation of the role of generative value co-creation. In contrast to a non-commercial platform running in the global South context, we observe that the focus is primarily on developing generative capacities of third-party developers (Msiska, 2018; Msiska et al., 2019). This observation might suggest the need for changes in governance of such commercial platforms if they are to support context specific challenges that affect those at the BoP.

# 6. CONCLUSION

The aim of this study was to examine to what extent commercially driven digital platforms facilitate ecosystem value co-creation in the global South using the case study of a mobile money platform. Although the findings suggest ecosystem value co-creation is enabled by introduction of boundary resources tools, there are impediments that are hindering third-party actors harnessing the generativity of the platforms. Taking all this together, we can conclude that platform generativity on its own is not sufficient to support ecosystem value co-creation but requires appropriate platform governance to deal with behavioural complexity of ecosystem actors using optimal control mechanisms.

## REFERENCES AND CITATIONS